\documentclass[twocolumn,english,aps,pra,floatfix,tightenlines,superscriptaddress]{revtex4}
\usepackage[T1]{fontenc}
\usepackage[latin9]{inputenc}
\usepackage{amsmath}
\usepackage{graphicx}
\usepackage{amssymb}
\usepackage{esint}

\makeatletter
\@ifundefined{textcolor}{}
{%
 \definecolor{BLACK}{gray}{0}
 \definecolor{WHITE}{gray}{1}
 \definecolor{RED}{rgb}{1,0,0}
 \definecolor{GREEN}{rgb}{0,1,0}
 \definecolor{BLUE}{rgb}{0,0,1}
 \definecolor{CYAN}{cmyk}{1,0,0,0}
 \definecolor{MAGENTA}{cmyk}{0,1,0,0}
 \definecolor{YELLOW}{cmyk}{0,0,1,0}
 }

\newcommand{\tr}{\mathrm{tr}}

\newcommand{\1}{\leavevmode{\rm 1\ifmmode\mkern  -4.8mu\else\kern -.3em\fi I}}

\usepackage{hyperref}

\usepackage{babel}
\usepackage{times}\usepackage{breakurl}

\usepackage{babel}

\makeatother

\usepackage{babel}

\begin{document}

\title{Quantification and Control of Non-Markovian Evolution in Finite Quantum
Systems via Feedback}

\author{N.~Chancellor}

\affiliation{Department of Physics and Astronomy and Center for Quantum Information
Science \& Technology, University of Southern California, Los Angeles,
CA 90089-0484, USA}

\affiliation{London Centre for Nanotechnology, University College London, London
WC1H 0AH, UK}

\author{C. Petri}

\affiliation{Department of Physics and Astronomy and Center for Quantum Information
Science \& Technology, University of Southern California, Los Angeles,
CA 90089-0484, USA}

\author{L.~Campos Venuti }

\affiliation{Department of Physics and Astronomy and Center for Quantum Information
Science \& Technology, University of Southern California, Los Angeles,
CA 90089-0484, USA}

\author{A.F.J.~Levi}

\affiliation{Department of Electrical Engineering, University of Southern California,
Los Angeles, CA 90089-2533, USA}

\affiliation{Department of Physics and Astronomy and Center for Quantum Information
Science \& Technology, University of Southern California, Los Angeles,
CA 90089-0484, USA}

\author{S.~Haas}

\affiliation{Department of Physics and Astronomy and Center for Quantum Information
Science \& Technology, University of Southern California, Los Angeles,
CA 90089-0484, USA}

\affiliation{School of Engineering and Science, Jacobs University Bremen, Bremen
28759, Germany}
\begin{abstract}
We consider the unitary time evolution of continuous quantum mechanical
systems confined to a cavity in contact with a finite bath of variable
size. We define a new measure relating to (non-)Markovianity which
parallels the standard one for the case of integrable Lindbladian
dynamics but has the advantage of being numerically tractable also
for large many particles systems\emph{.} The relevant time scales
are identified, which characterize non-Markovian transient behavior,
boundary scattering induced non-Markovian oscillations at intermediate
times, and non-Markovian rephasing events at long time scales. It
is shown how these time scales can be controlled by tunable parameters
such as the bath size and the strength of the system-bath coupling. 
\end{abstract}
\maketitle

\section{Introduction}

It is well known that the time evolution of quantum closed systems
at finite size is intrinsically non-Markovian. The reason is that
the reduced density matrix is necessarily an oscillating function
which admits no infinite-time limit. In the quantum case, the time
scale for partial rephasing events (revivals) is proportional to the
system size, whereas full rephasing (Poincare recurrences) can occur
at an astronomically large time scales \cite{campos_venuti_universality_2010,campos_venuti_equilibration_2013}.
Nonetheless, previous studies of the transient dynamics in lattice
models have revealed extended time domains within which the time evolution
is pseudo-Markovian in the sense that memory of the initial state
appears to be lost \cite{chancellor_non-markovian_2013}. Strictly
speaking, this conclusion would of course be erroneous, as this information
is only temporarily dispersed among the system's accessible modes
before it re-emerges during rephasing events. However, this observation
raises the interesting question if and how the time scales which determine
such pseudo-Markovian time evolution (i.e.~Markovian when restricted
to a definite time domain) as well as non-Markovian features can be
controlled.

Here we consider the illustrative example of a continuous one-dimensional
array with periodic boundary conditions. We divide this array into
an active {}``system\textquotedblright{} in region $A$ connected
to a region $B$ which acts as a bath. A pair of thin barriers are
than placed between the system and the bath. The concept of a finite
bath is different from the conventional notion of an infinite reservoir,
but this setup allows us to explore the size of the bath as a non-trivial
tuning parameter, along with the system-bath coupling. In previous
work, the non-Markovian dynamics in such arrangements have been explored
using lattice Hamiltonians \cite{chancellor_non-markovian_2013}.
It was found that by breaking certain symmetries, features signaling
non-Markovian time evolution disappear within a finite time scale
which is governed by the system-bath coupling as well as by the amount
of random symmetry breaking introduced in the bath Hamiltonian. Here,
randomness raises the number of accessible states of the bath.

In this paper, we investigate control over the time evolution of continuum
Hamiltonians, tuning their bath degrees of freedom by adjusting the
bath size relative to the system. While this study is kept at a relatively
abstract level, primarily focusing on the emerging time scales, experimental
realizations of such configurations can easily be imagined in the
context of coupled laser cavities, connected by semi-transparent mirrors,
or in the context of electron wave packets tunneling through barriers
in layered (Ga,Al)As heterostructures. In order to avoid the pitfalls
of a more conventional treatment of open quantum systems based on
the Lindblad formalism \cite{lindblad_generators_1976}, including
infinitesimal system-bath coupling and forced Markovianity, we study
the time evolution of a complete, untruncated system-bath Hamiltonian.
The price we have to pay for this is the restriction to non-interacting
systems described by effective single-particle Hamiltonians, in order
to keep the problem numerically manageable.

\begin{figure}
\begin{centering}
\includegraphics[width=6cm]{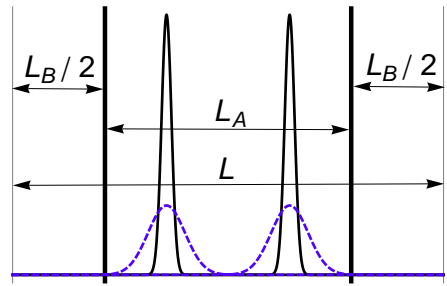} 
\par\end{centering}

\caption{(Color online) Illustration of the finite system-bath array considered
here. The system-bath boundaries are shown as thick vertical lines,
along with two sets of initial Gaussian wave functions: broad (dashed
line) and narrow (continuous line). Except otherwise stated, delta
barriers on the left system-bath boundary are of strength $V_{1}=10^{6}$
and on the right boundary are of strength $V_{2}=2.0\times10^{6}$.
The bath segments are connected via periodic boundary conditions.
\label{fig:system_bath}}
\end{figure}

We will first introduce the concept of free Markovianity (FM) which
parallels the definition of Markovianity a-la Breuer-Laine-Piilo (BLP)
\cite{breuer_measure_2009}, for the case in which the Lindblad, Markovian
dynamics, is integrable in the sense of \cite{prosen_third_2008}.
Free Markovianity differs from BLP Markovianity in the general case,
however the two definitions become equivalent when restricted to integrable
dynamics. The resulting measure of free non-Markovianity (FNM), $D(t)$,
detects a free non-Markovian event whenever $D(t)$ is an increasing
function of time, in analogy with the BLP measure \cite{breuer_measure_2009}.
It should be reminded at this point that properly quantifying the
amount of non-Markovianity of a given dynamics has been subject of
intense research in recent times and many different, non-necessarily
equivalent, measures of non-Markovianity have been proposed \cite{benenti_enhancement_2009,breuer_measure_2009,rivas_entanglement_2010,mazzola_dynamical_2012,luo_quantifying_2012,bylicka_non-markovianity_2013}.
The measure $D(t)$ that we propose, although identifies yet another
class of non-Markovian systems, has a series of advantages with respect
to the measures that have appeared so far. The main point being that
$D(t)$ can easily be computed even for many-body systems composed
of many particles%
\footnote{As will be clear below, the computation simply requires diagonalization
of a matrix of size $O(N)$ for a system of $N$ particles.%
}. Moreover the quantity $D(t)$ is expressed in terms of single particles
quantities thus allowing for a simple physical interpretation. While
in general, the maximum of $D(t)$ over all possible initial states
has to be calculated, even a single pair of linearly independent states
can provide valuable insight into the overall dynamics of the system.
We will refer to the distance between such a pair as $D_{\Gamma_{1},\Gamma_{2}}(t)$,
where $\Gamma$ are the covariance matrix of the two states. 

This paper is organized as follows. In the next section we define
our measure of free non-Markovianity for non-interacting systems.
We then describe the general framework. This is followed by an analysis
of the characteristic time-scales that can be extracted from the numerics.
We conclude with a discussion of how feedback via tunable bath parameters
can be utilized to control the time evolution in physically relevant
systems.

\section{Measure of free non-Markovianity for non-interacting systems\label{sec:Measure-of-non-Markovianity}}

We proceed here to define a measure of free non-Markovianity for a
system of non-interacting particles. The results apply both to Fermi-Dirac
and Bose-Einstein statistics although with some technical caveats
in the latter case. The proper definition of a measure of non-Markovianity
is an issue of current debate \cite{breuer_measure_2009,chruscinski_non-markovian_2010,rivas_entanglement_2010,chruscinski_measures_2011,bylicka_non-markovianity_2013}.
For our purposes it seems appropriate to take the point of view of
Breuer-Laine-Piilo (BLP) \cite{breuer_measure_2009} which provides
a physical interpretation in terms of information flow. According
to \cite{breuer_measure_2009} a violation of Markovianity is detected
whenever the quantity $\left\Vert \rho_{1}(t)-\rho_{2}(t)\right\Vert _{1}$
increases in time. Here $\left\Vert \cdot\right\Vert _{1}$ denotes
the trace norm and $\rho_{i}(t)$ are the density matrices of the
system of interest at time $t$, corresponding to different initial
conditions $\rho_{1}(0),\,\rho_{2}(0)$. As it turns out the trace
norm $\left\Vert \rho_{1}(t)-\rho_{2}(t)\right\Vert _{1}$ is difficult
to compute even for the simplest case of systems composed of a single
qubit \cite{breuer_measure_2009}. Therefore we take a different approach.

Consider a many-body system whose dynamical evolution is not necessarily
unitary, but it is assumed to be non-interacting in a sense that we
are going to specify shortly. Suppose that the quantum process is
described by a Markovian master equation,\begin{equation}
\frac{d\rho}{dt}=\mathcal{L}\rho\label{eq:Lindblad}\end{equation}
 with generator $\mathcal{L}$ in Lindblad form \begin{equation}
\mathcal{L}\rho=-i\left[H,\rho\right]+\sum_{i}\gamma_{i}\left[A_{i}\rho A_{i}^{\dagger}-\frac{1}{2}\left\{ A_{i}^{\dagger}A_{i},\rho\right\} \right].\end{equation}
In the spirit of \cite{prosen_third_2008} we call the dynamics non-interacting
if the Hamiltonian $H$ is quadratic in the canonical creation and
annihilation operators whereas the Lindblad terms $A_{i}$ are linear
combinations thereof. The solution of such a free Markovian master
equations is conveniently encoded in terms of the covariance matrix
or two-point correlation function $\Gamma_{j\alpha,k\beta}$ \cite{prosen_quantization_2010,prosen_spectral_2010,eisert_noise-driven_2010}.
For Fermions the covariance matrix has the form $\Gamma_{j\alpha,k\beta}=-\mathrm{Im[}\tr(\rho\omega_{j}^{\alpha}\omega_{k}^{\beta})${]}
where the the Majorana operators $\omega_{j}^{\alpha}$ are given
by $\omega_{j}^{1}=f_{j}+f_{j}^{\dagger}$, $\omega_{j}^{2}=i(f_{j}-f_{j}^{\dagger})$
in terms of Fermi operators $f_{i}$. For bosons instead one has $\Gamma_{j\alpha,k\beta}=\mathrm{Re[}\tr(\rho u_{j}^{\alpha}u_{k}^{\beta})${]}
for quadrature operators $u_{j}^{1}=b_{j}+b_{j}^{\dagger}$ and $u_{j}^{2}=i(b_{j}-b_{j}^{\dagger})$
and canonical Bose operators $b_{j}$ (see Ref.~\cite{eisert_noise-driven_2010}
for more details). 

From Equation (\ref{eq:Lindblad}) one finds that the covariance matrix
$\Gamma$ satisfies the equation of motion \cite{eisert_noise-driven_2010}\begin{equation}
\frac{d\Gamma}{dt}=X^{T}\Gamma+\Gamma X-Y,\label{eq:Lyap}\end{equation}
 with matrices $X,Y$ which depend on $H$ and $A_{i}$. In the Fermionic
case one can show that the spectrum of $X$ lies in the sector $\mathrm{Re}\left(z\right)\ge0$
of the complex plane \cite{prosen_quantization_2010,eisert_noise-driven_2010}.
This in turns implies that the dynamics given by Eq.~(\ref{eq:Lindblad})
gives rise to a contractive flow. In other words, using following
basis-dependent identification between matrices and vectors $|\,|n\rangle\langle m|\gg:=|n,m\rangle$
(see e.g.~\cite{watrous_lecture_2011} Sec.~2.4) given two different
initial conditions $\Gamma_{1}(0),\,\Gamma_{2}(0)$, one has%
\footnote{In vector notation Eq.~(\ref{eq:Lyap}) becomes $|\dot{\Gamma}\gg=\left[\1\otimes X^{T}+X^{T}\otimes\1\right]|\Gamma\gg-|Y\gg$.
Calling $\mathcal{M}=\left[\1\otimes X^{T}+X^{T}\otimes\1\right]$
the solution is $|\Gamma(t)\gg=e^{t\mathcal{M}}\left(|\Gamma(0)\gg-|\Gamma(\infty)\gg\right)+|\Gamma(\infty)\gg$
with unique asymptotic state $|\Gamma(\infty)\gg=\mathcal{M}^{-1}|Y\gg$
in case $\mathcal{M}$ is invertible. Under the above hypothesis the
spectrum of $\mathcal{M}$ has non positive real part and the result
(\ref{eq:contractive}) follows. %
}\begin{equation}
\left\Vert |\Gamma_{1}(t)\gg-|\Gamma_{2}(t)\gg\right\Vert \le\left\Vert |\Gamma_{1}(s)\gg-|\Gamma_{2}(s)\gg\right\Vert ,\,\mathrm{for}\, s\le t.\label{eq:contractive}\end{equation}
 The same result does not apply directly in the case of bosons essentially
because the mapping $\rho\to\Gamma\left[\rho\right]$ is not continuous.
Physically this corresponds to the possibility of pumping-in infinite
energy in the bosonic fields \cite{prosen_quantization_2010,eisert_noise-driven_2010}.
A standard procedure to avoid such infinities is to introduce a (very
large) cut-off in the number of particle thus avoiding infinite energy
states. With this prescription Eq.~(\ref{eq:contractive}) holds
also for bosons. 

The norm appearing in Eq.~(\ref{eq:contractive}) is in principle
any norm for the vectors $|\Gamma_{j}(t)\gg$. The $\ell^{2}$ norm
seems to be the most natural one which induces a basis-independent
norm on the matrices $\Gamma_{j}(t)$. The induced norm in this case
is the Hilbert-Schmidt (HS) norm for the matrices $\Gamma_{j}$, and
one has $\left\Vert |X\gg\right\Vert _{\ell^{2}}=\left\Vert X\right\Vert _{HS}=\sqrt{\tr\left(X^{\dagger}X\right)}$.

To summarize, the BLP measure of Markovianity \cite{breuer_measure_2009}
is defined as the trace distance between two density matrices,

\begin{equation}
D_{\rho_{1},\rho_{2}}^{BLP}(t)=\left\Vert \rho_{1}(t)-\rho_{2}(t)\right\Vert _{1},\end{equation}

while our measure of \emph{free Markovianity} is defined as the Hilbert-Schmidt
distance between two covariance matrices

\begin{equation}
D_{\Gamma_{1},\Gamma_{2}}(t)=\left\Vert \Gamma_{1}(t)-\Gamma_{2}(t)\right\Vert _{HS}.\end{equation}

In the spirit of Ref.~\cite{breuer_measure_2009} we now single out
a violation of \emph{free Markovianity} if, for some initial states
$\Gamma_{1}(0),\,\Gamma_{2}(0)$, the norm $\left\Vert \Gamma_{1}(t)-\Gamma_{2}(t)\right\Vert _{HS}$
is not decreasing. In other words, calling $\sigma\left(t,\Gamma_{1},\Gamma_{2}\right):=(d/dt)\, D_{\Gamma_{1},\Gamma_{2}}(t)=(d/dt)(\left\Vert \Gamma_{1}(t)-\Gamma_{2}(t)\right\Vert _{HS}/2)$,
we say that the dynamics is free Markovian (FM) if for some time interval
and initial states with covariance matrix $\Gamma_{1,2}(0)$, we have
$\sigma\left(t,\Gamma_{1},\Gamma_{2}\right)\le0$. In principle one
could even define a measure of free non-Markovianity paralleling the
definition of \cite{breuer_measure_2009} for quasi-free dynamics.
The quantity:\begin{equation}
\mathcal{N}=\sup_{\Gamma_{1},\Gamma_{2}}\int_{\sigma>0}\sigma(t,\Gamma_{1},\Gamma_{2})dt,\label{eq:total_NM}\end{equation}
 encodes the amount of free non-Markovianity in the process dynamics
from $t=0$ to $t=\infty$. As we will see, for one-particle states
$\left\Vert \Gamma_{1}(t)-\Gamma_{2}(t)\right\Vert _{HS}$ has a particularly
simple form so that one can hope to be able to perform the maximization
in Eq.~(\ref{eq:total_NM}). This project is left for future investigations.

In general free Markovianity is different from the notion of Markovianity
a-la BLP. However, when restricted to the class of free, Lindblad
dynamics the two definitions become equivalent. 

Our procedure will be the following. We will consider a system of
identical particles described by a free Hamiltonian dynamics. The
setting is relevant to electrons, so that particles will be fermions,
but, as shown above, the statistic is essentially unimportant modulo
a technical caveat. Restricting to free Hamiltonians will allow us
to easily integrate the equation of motions going to the one particle
sector. We then consider the sub-dynamics of the system, obtained
by tracing over the bath degrees of freedom. The central question
we ask, is if such sub-dynamics can be described by a free-Markovian
master equation. To this end we compute the quantity in Eq.~(\ref{eq:contractive})
for different initial states and Hamiltonian parameters. Note that
the result, Eq.~(\ref{eq:contractive}), is expressed in terms of
single-particle quantities. This fact will make the physical interpretation
simpler. In the following, for simplicity of language, we will simply
refer to free (non) Markovianity as (non) Markovianity. The definition
given here is intended throughout unless otherwise specified.

\section{Methods}

We now explain the general setting. We place a single particle in
a one-dimensional array of length $L=L_{A}+L_{B}$, with periodic
boundary conditions (PBC), i.e.~a ring. We will later trace out the
segment $B$ of length $L_{B}$ which plays the role of external bath,
whereas $A$ is the system of interest. The one-particle Hamiltonian
is of the form $H=-\partial^{2}/\partial x^{2}+V(x)$ where the external
potential $V(x)$ will be specified later. In a particle-number conserving
system the covariance matrix $\Gamma$ is a function of $R_{x,y}:=\tr(\rho c_{x}^{\dagger}c_{y})$
only and one has $\left\Vert \Gamma\right\Vert _{HS}=\sqrt{8}\left\Vert R\right\Vert _{HS}$
\footnote{We are aware that a particle-number conserving evolution is not consistent
with Eqns.~(\ref{eq:Lindblad}) and (\ref{eq:Lyap}). In practice
we are asking if it is possible to have a Markovian evolution {[}of
the form of Eqns.~(\ref{eq:Lindblad}) and (\ref{eq:Lyap}){]} for
the subsystem $A$ giving rise to the observed matrix $R$ in this
subsystem. %
}. The system is initialized in the one-particle state $|\psi(0)\rangle$,
which is then evolved according to $|\psi(t)\rangle=e^{-itH}|\psi(0)\rangle$.
In the Fock space this corresponds to the state $|\Psi(t)\rangle=c_{\psi(t)}^{\dagger}|0\rangle$
where $c_{f}^{\dagger}$ create a Fermion or Boson in state $f$.
Now, in the position {}``basis'' $|x\rangle$, $|\psi(t)\rangle=\sum_{x}\psi(x,t)|x\rangle$
and $c_{\psi(t)}^{\dagger}=\sum_{x}\overline{\psi(x,t)}c_{x}^{\dagger}$
and one obtains the following integral kernel \begin{eqnarray}
R_{x,y} & = & \overline{\psi(x,t)}\psi(y,t).\label{eq:cov-1P}\end{eqnarray}
The full density matrix is a many body state $\rho(t)=|\Psi(t)\rangle\langle\Psi(t)|$,
and tracing out the bath $B$ corresponds to discarding from $R$
in Eq.~(\ref{eq:cov-1P}) all the labels $(x,y)$ belonging to $B$,
i.e.~projecting $R$ onto $A$. The one-particle Hilbert space is
decomposed into a direct sum $\mathcal{H}=\mathcal{H}_{A}\oplus\mathcal{H}_{B}$,
where $\mathcal{H}_{A/B}$ describes wave functions with support only
in $A/B$. Calling $P_{A}$ the operator which projects onto $\mathcal{H}_{A}$,
the restriction of $R=|\psi\rangle\langle\psi|$ on $\mathcal{H}_{A}$
is $R^{A}=P_{A}|\psi\rangle\langle\psi|P_{A}=|\psi_{A}\rangle\langle\psi_{A}|$,
for a non-normalized state $|\psi_{A}\rangle$. With the notation
$\rho_{R}$ for the Gaussian state given by covariance $R$, the tracing
out the bath is simply achieved via  $\tr_{B}\rho_{R}=\rho_{R^{A}}$.
In the general case in which the initial states has $N$ particles,
$R$ and $R^{A}$ are rank $N$ operators. Considering the evolution
of two different initial states $|\psi_{j}(0)\rangle$, the difference
of covariance matrices restricted to region $A$ is \begin{equation}
P_{A}(|\psi_{1}\rangle\langle\psi_{1}|-|\psi_{2}\rangle\langle\psi_{2}|)P_{A}=|\psi_{1,A}\rangle\langle\psi_{1,A}|-|\psi_{2,A}\rangle\langle\psi_{2,A}|\end{equation}
 with unnormalized states $|\psi_{j,A}\rangle$ supported in $\mathcal{H}_{A}$.
In the, non-orthonormal, basis $|\psi_{j,A}\rangle$ the above operator
has the form \begin{equation}
\left(\begin{array}{cc}
p_{1,1} & p_{1,2}\\
-\overline{p_{1,2}} & -p_{2,2}\end{array}\right),\label{eq:matrix}\end{equation}
 where \begin{eqnarray}
p_{i,j} & = & \langle\psi_{i,A}|\psi_{j,A}\rangle\\
 & = & \langle\psi_{i}(t)|P_{A}|\psi_{j}(t)\rangle\\
 & = & \sum_{k,q}e^{it(E_{k}-E_{q})}\langle\psi_{i}|\phi_{k}\rangle\langle\phi_{q}|\psi_{j}\rangle\Delta_{L}^{A}(k,q)\label{eq:pij_def}\end{eqnarray}
 with $i,j=1,2$ and having defined\begin{equation}
\Delta_{L}^{A}(k,q)=\langle\phi_{k}|P_{A}|\phi_{q}\rangle=\int_{A}\overline{\phi_{k}(x)}\phi_{q}(x)dx.\label{eq:kernel_A}\end{equation}
 The term $p_{j,j}$ gives the probability that the particle initialized
in $|\psi_{j}\rangle$ is in region $A$ at time $t$. Our indicators
of non-Markovianity are given in terms of the eigenvalues of the matrix
(\ref{eq:matrix}) which read

\begin{equation}
\lambda_{1,2}=\frac{\left(p_{1,1}-p_{2,2}\right)\pm\sqrt{\left(p_{1,1}+p_{2,2}\right)^{2}-4\left|p_{1,2}\right|^{2}}}{2}.\label{eq:lambda_1_2}\end{equation}

Finally, according to the discussion in Sec.~\ref{sec:Measure-of-non-Markovianity},
the distance that we consider to characterize non-Markovianity is
given by \begin{align}
D_{\psi_{1},\psi_{2}} & =\frac{1}{\sqrt{2}}\left\Vert R_{1}^{A}(t)-R_{2}^{A}(t)\right\Vert =\frac{1}{\sqrt{2}}\sqrt{\lambda_{1}^{2}+\lambda_{2}^{2}}\\
 & =\frac{1}{\sqrt{2}}\sqrt{p_{1,1}^{2}+p_{2,2}^{2}-2\left|p_{1,2}\right|^{2}},\label{eq:D_Markov}\end{align}
where $R_{j}^{A}(t)$ are the covariance matrices of system $A$ at
time $t$. A factor $1/\sqrt{2}$ has been inserted to scale the measure,
so that its maximum value is $1$ (attained when $p_{1,1}=p_{2,2}=1$
and $p_{1,2}=0$). Note that by Schwartz inequality one has $\left|p_{1,2}\right|^{2}\le p_{1,1}p_{2,2}$
implying that $D_{\psi_{1},\psi_{2}}$ is indeed real.

The final result Eq.~(\ref{eq:D_Markov}) is extremely simple and
physically quite compelling. Assume for simplicity that the states
are orthogonal on $A$ (so that $p_{1,2}=0$). Eq.~(\ref{eq:D_Markov})
then is simply the geometric mean of the probabilities of the particles
being in region $A$. As such, as a function of $t$, it is quite
clear that $D_{\psi_{1},\psi_{2}}(t)$ increases when particles move
into region $A$ signaling a violation of Markovianity. Since $D_{\psi_{1},\psi_{2}}(t)$
is the (Hilbert-Schmidt) distance of two covariance matrices it trivially
characterizes the distinguishability of the $R_{j}^{A}(t)$. The {}``information
flow'' of the BLP measure becomes in this setting a flow of probability.
With slight abuse of language we will speak of Markovian behavior
when $D_{\psi_{1},\psi_{2}}(t)$ decreases in time although this is
only consistent with Markovian dynamics. This is in accordance with
the general intuition of Markovian evolution occurring due to information
leakage from the system.

\section{Results}

We consider two specific initial Gaussian packets localized around
$x_{j}$: $\psi_{j}(x)=C_{j}e^{-(x-x_{j})^{2}/(4\sigma_{j}^{2})}$
($j=1,2$). The normalization factor is given by the equation $C_{j}^{2}\sqrt{\pi/2}\sigma_{j}\{\mathrm{Erf}\left[(L-2x_{j})/(\sqrt{8}\sigma_{j})\right]$
$+\mathrm{Erf}\left[(L+2x_{j})/(\sqrt{8}\sigma_{j})\right]\}=1$,
\footnote{The error function is defined as $\mathrm{Erf}(z):=\pi^{-1/2}2\int_{0}^{z}e^{-t^{2}}dt$.%
} which boils down to $C_{j}\simeq\left(2\pi\sigma_{j}^{2}\right)^{-1/4}$
for $\sigma_{j}\ll L$. We choose initial states symmetrically displaced
with respect to the origin, i.e.~$x_{1}=-x_{2}$ as shown in Fig.~\ref{fig:system_bath}.
In this case, for symmetric potentials $V(x)$ one has $p_{1,1}(t)=p_{2,2}(t)$
\footnote{Indeed, consider $U_{LR}$ the unitary operator which implements the
Left-Right inversion around the origin. Then, at all times, for symmetric
potentials $V(x)$, $U_{LR}|\psi_{1}(t)\rangle=e^{i\phi}|\psi_{2}(t)\rangle$.
Then $p_{1,1}(t)=\langle\psi_{1}(t)|U_{LR}^{\dagger}U_{LR}P_{A}U_{LR}^{\dagger}U_{LR}|\psi_{1}(t)\rangle=\langle\psi_{2}(t)|P_{A}|\psi_{2}(t)\rangle=p_{2,2}(t)$
since $[P_{A},U_{LR}]=0$.%
}. For such symmetric configurations $\lambda_{1}=-\lambda_{2}$, and
one has the further simplification $D_{\psi_{1},\psi_{2}}=\left|\lambda_{1}\right|=\left|\lambda_{2}\right|$.
As a first example we consider in some detail the purely kinetic evolution
corresponding to $V(x)=0$.

\subsection{Purely kinetic evolution}

Let us first consider the case when there are no delta barriers separating
the system from the bath, as shown in the insets (a) and (c) of Fig.~\ref{fig:Time-evolution}.
The spectrum acquires the familiar form $E_{k}=k^{2}$ where the quasi-momenta
satisfy $k=2\pi n/L$, $n\in\mathbb{Z}$. The {}``geometric'' factor
$\Delta_{L}^{A}$ becomes \begin{equation}
\Delta_{L}^{A}(k,q)=\frac{1}{L}\int_{L_{A}}e^{ix(k-q)}dx=\frac{\sin\left[L_{A}(k-q)/2\right]}{\left[L(k-q)/2\right]}.\end{equation}

\begin{figure}
\begin{centering}
\includegraphics[width=4cm,height=3cm]{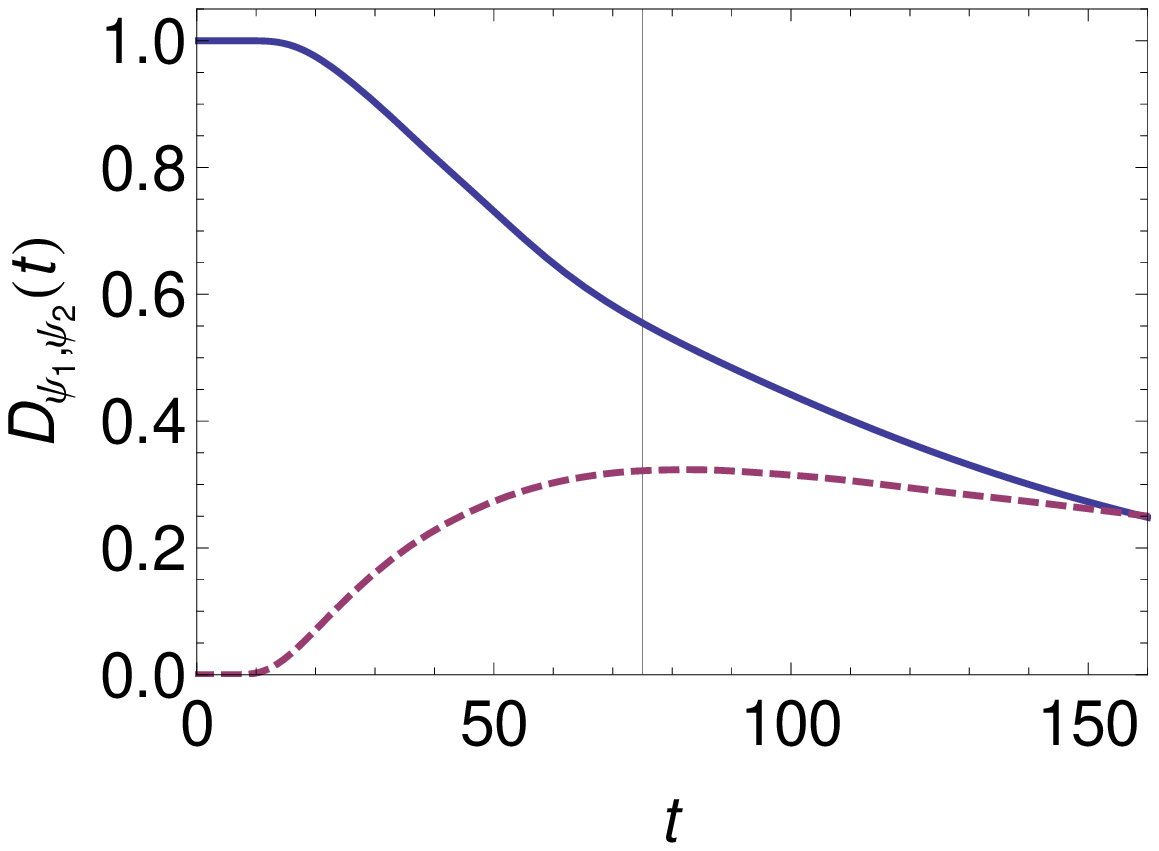} \includegraphics[width=4cm,height=3cm]{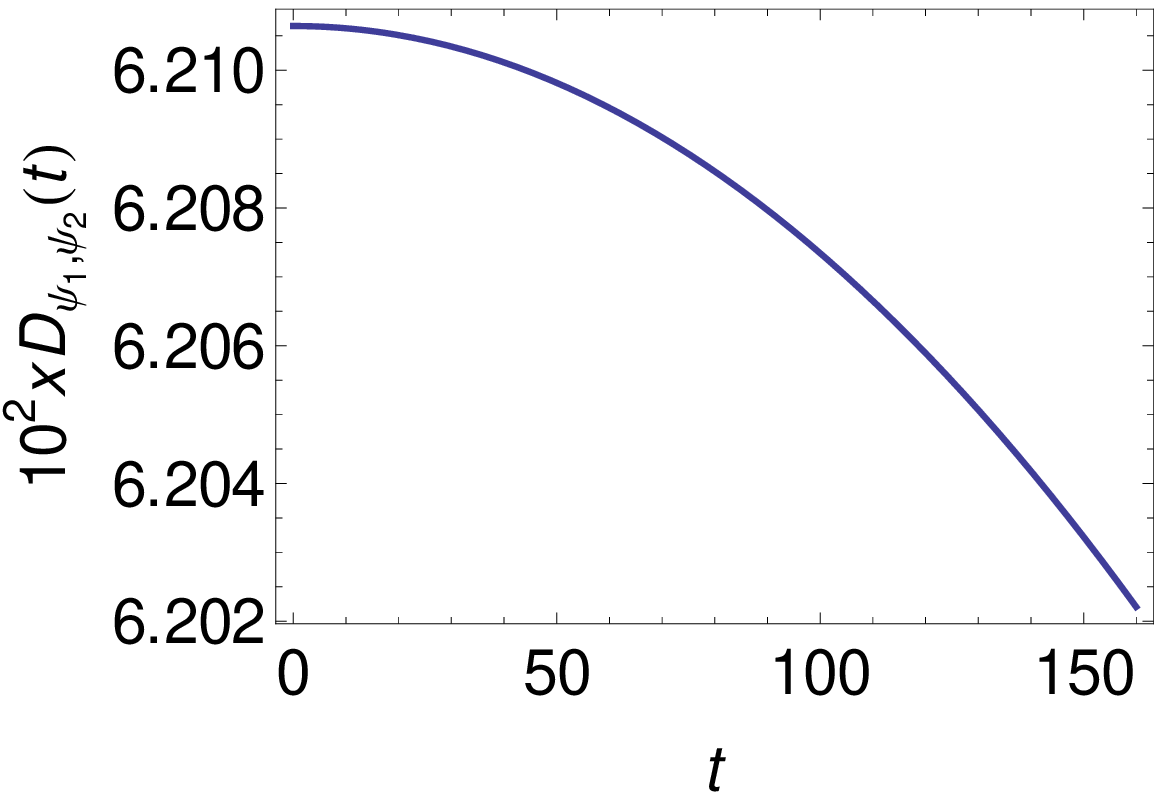}
\par\end{centering}

\caption{(Color online) Markovian/non-Markovian character for an infinite bath.
Left panel: continuous curve for initial states {}``inside region
$A$'' ($x_{1}=-x_{2}=10$, $L_{A}=60$ $\sigma=2$), dashed curve,
initial states outside region $A$ ($x_{1}=-x_{2}=45$, $L_{A}=60$
$\sigma=2$). Here a non-Markovian window appears for $0\le t\le\tau_{M}$
with $\tau_{M}\approx\sigma\,(x_{1}+L_{A}/2)/2$ indicated by the
vertical line. Note that $D_{\psi_{1},\psi_{2}}(t)$ starts decreasing
roughly for $t\ge\tau_{M}$. Right panel: if the initial width $\sigma$
is cranked beyond a value proportional to $(x_{1}+L_{A}/2)/2$ the
non-Markovian window disappears. Here $x_{1}=-x_{2}=45$, $L_{A}=60$
$\sigma=60$. Note the very small vertical scale, i.e.~$D_{\psi_{1},\psi_{2}}(t)$
is almost constant. \label{fig:Markovian-behavior}}
\end{figure}

In this case we have simple dispersing wave packets. However, since
the system-bath configuration has finite spatial extent, there are
rephasing events (or revivals) that occur at time scales of the order
of the total system size, i.e. $\tau_{L}\propto L$ \cite{campos_venuti_unitary_2010,happola_universality_2012}.
This timescale measures the time it takes for a packet to go around
the periodic boundaries and come back. As Fig. \ref{fig:Markovian-behavior}
shows, recurrences are not present if the bath is infinite (discussed
in more detail later). However, for a finite bath recurrences are
observed in Fig.~\ref{fig:Time-evolution} (a) and (c), where the
time evolution of the non-Markovian indicator between two initial
wave packets is shown. In addition to these rephasing events one also
observes smaller amplitude, more rapid oscillations in the non-Markovian
indicator. A measure of these oscillations is roughly given by the
effective numbers of Hamiltonian eigenstates needed to express the
initial wave-packet. One can obtain this number by imposing that the
fidelity of the initial state be $F=1-\epsilon$. One obtains that
roughly $N^{\ast}=-(L/\sigma)\sqrt{-\ln\left(\epsilon\right)}$ eigenstates
are required to obtain the desired fidelity. The number $N^{\ast}$
also gives the number of effective energy eigenstates involved in
the dynamics. It is evident from the figure that broad initial wave-packets
(Fig.~\ref{fig:Time-evolution} (c) and (d)) contain many more frequencies
than narrow initial wave-packets (Fig.~\ref{fig:Time-evolution}
(a) and (b)). In fact smaller $\sigma$, means larger $N^{\ast}$
so that the measure $D_{\psi_{1},\psi_{2}}(t)$ contains more frequencies
and consequently shows faster oscillations.

We conclude that for this most simple example non-Markovian time evolution,
i.e.~deviations from a monotonically decaying HS distance, occurs
at two time scales, $\tau_{L}$ controlled by the total system size,
and $\tau_{\sigma}$ controlled by the width of the initial wave packets
and their subsequent dispersion.

We consider now the situation where $L\to\infty$. In this case many
time scales which depend on $L$, such as $\tau_{L}$, are sent to
infinity. We then keep $L_{A}$ constant and send $L\to\infty$ in
Eq.~(\ref{eq:pij_def}). In this limit $L^{-1}\sum_{k}\to(2\pi)^{-1}\int_{\mathbb{R}}dk$,
and we obtain \begin{equation}
p_{i,j}=\int_{\mathbb{R}}\frac{dk}{2\pi}\int_{\mathbb{R}}\frac{dq}{2\pi}\, e^{it(k^{2}-q^{2})}\overline{\hat{\psi}_{i}(k)}\hat{\psi}_{j}(q)\frac{\sin\left[L_{A}(k-q)/2\right]}{\left[(k-q)/2\right]},\label{eq:pij-free}\end{equation}
 where the initial wave functions in Fourier space are given by \begin{equation}
\hat{\psi}_{j}(q)=\int_{\mathbb{R}}dxe^{iqx}\psi_{j}(x)=\sqrt{4\pi\sigma_{j}^{2}}C_{j}e^{iqx_{j}}e^{-q^{2}\sigma_{j}^{2}}.\end{equation}

Changing variables to $r=(k-q),\, R=(k+q)/2$, the integral over $R$
is Gaussian, and we obtain \begin{align}
p_{i,j} & =\int_{\mathbb{R}}dr\,\frac{\sin(rL_{A}/2)}{\pi r}\exp\left\{ -\frac{r^{2}\sigma_{t}^{2}}{2}\right.\\
 & \left.+\frac{r}{2}\left[\frac{t}{\sigma^{2}}(x_{i}-x_{j})-i\left(x_{i}+x_{j}\right)\right]-\frac{(x_{i}-x_{j})^{2}}{8\sigma^{2}}\right\} ,\end{align}
 with$\sigma_{t}=\sqrt{\left(t^{2}/\sigma^{2}+\sigma^{2}\right)}$.
Writing $\sin\left(L_{A}r\right)/r=\int_{0}^{L_{A}}\cos\left(yr\right)dy$,
the integral over $r$ is Gaussian, and one is left with an incomplete
Gaussian integral over $y$. The final result is \begin{multline}
p_{i,j}=\frac{e^{-(x_{i}-x_{j})^{2}/(8\sigma^{2})}}{2}\times\\
\left[\mathrm{Erf}\left(\frac{L_{A}^{+}\sigma^{2}+it(x_{i}-x_{j})}{\sqrt{8}\sigma_{t}\sigma^{2}}\right)\right.\\
\left.+\mathrm{Erf}\left(\frac{L_{A}^{-}\sigma^{2}-it(x_{i}-x_{j})}{\sqrt{8}\sigma_{t}\sigma^{2}}\right)\right]\label{eq:pij-final-1}\end{multline}
 having defined $L_{A}^{\pm}=L_{A}\pm(x_{i}+x_{j})$. Plugging Eq.~(\ref{eq:pij-final-1})
into Eq.~(\ref{eq:D_Markov}) one obtains\begin{eqnarray}
\left(D_{\psi_{1},\psi_{2}}\right)^{2} & \!\!=\!\! & \frac{1}{8}\sum_{j=1}^{2}\left[\mathrm{Erf}\left(\frac{L_{A}-2x_{j}}{\sqrt{8}\sigma_{t}}\right)+\mathrm{Erf}\left(\frac{L_{A}+2x_{j}}{\sqrt{8}\sigma_{t}}\right)\right]^{2}\nonumber \\
 & - & \frac{1}{4}e^{-\frac{(x_{1}-x_{2})^{2}}{2\sigma^{2}}}\left|\mathrm{Erf}\left(\frac{L_{A}^{+}}{\sqrt{8}\sigma_{t}}+it\frac{(x_{1}-x_{2})}{\sqrt{8}\sigma_{t}\sigma^{2}}\right)\right.\nonumber \\
 &  & \left.+\mathrm{Erf}\left(\frac{L_{A}^{-}}{\sqrt{8}\sigma_{t}}-it\frac{(x_{1}-x_{2})}{\sqrt{8}\sigma_{t}\sigma^{2}}\right)\right|^{2}\label{eq:D2-final}\end{eqnarray}
 For simplicity, as stated already, we initiate the evolution in a
symmetric configuration where $x_{2}=-x_{1}$. To discuss the Markovian
character of the evolution, encoded in the above equation we have
to distinguish two cases according to whether the initial wave packets
are centered inside or outside region $A$ (i.e.~$\left|x_{i}\right|\le L_{A}/2$
or $\left|x_{i}\right|>L_{A}/2$). Intuitively, in the first case
particles can only escape from region $A$, and so we always expect
Markovianity for any parameter value. Indeed this intuition can be
confirmed after a lengthy calculation taking the time derivative of
Eq.~(\ref{eq:D2-final}).

Let us now consider the other situation where the particles are initialized
outside $A$, i.e.~$\left|x_{i}\right|>L_{A}/2$. In this case, the
wave front first enters region $A$ after having traveled a distance
$x_{1}-L_{A}/2$, then travels inside $A$ (length $L_{A}$), and
finally escapes region $A$ after having traveled a length $x_{1}+L_{A}/2$.
At the beginning, particles enter region $A$, and we expect an increase
in $D_{\psi_{1},\psi_{2}}(t)$. Assuming that the front moves at constant
speed $v\sim2/\sigma$ (which can be read off from $\sigma_{t}$ at
large times) the time scales after which we expect to see Markovian
behavior is given roughly by $\tau_{M}\simeq\sigma\,(x_{1}+L_{A}/2)/2$.
In general, the appearance of a region of non-Markovianity can be
observed as long as the initial wave functions are sufficiently localized,
so that the initial front does not surpass region $A$, i.e.~roughly
$\sigma<(x_{1}+L_{A}/2)/2$.

These predictions are confirmed by our numerical experiments shown
in Fig.~\ref{fig:Markovian-behavior}. The purely Markovian behavior
observed for $\left|x_{i}\right|\le L_{A}/2$ indicates, as it is
reasonable to expect, that for infinite spatial extent, wave packets
always leak out of region $A$ and never come back. An important ingredient
in reaching this conclusion is the fact that, for the case considered,
the spectrum is purely continuous. The presence of bound states in
the spectrum may lead to oscillations of information back and forth
from region $A$. This in turn may lead to a breaking of Markovianity
in case such bound states are initially populated.

\subsection{Double delta barrier}

\begin{figure}
\begin{centering}
\includegraphics[width=8cm,height=6cm]{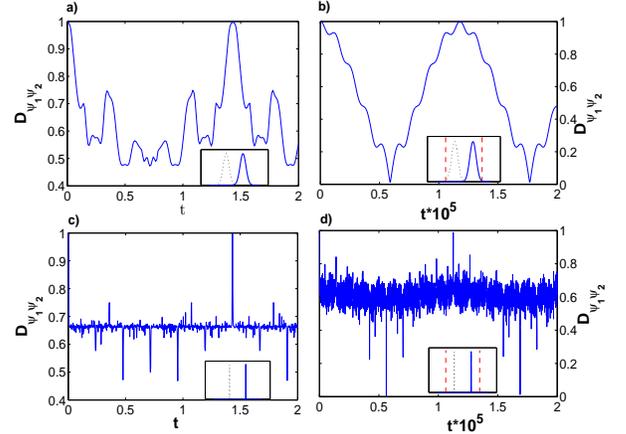} 
\par\end{centering}

\caption{(Color online) Time evolution of the HS distance $D_{\psi_{1},\psi_{2}}(t)$
between two Gaussian wave packets for various initial conditions with
and without system-bath delta barriers. The insets show initial conditions
and whether or not barriers are included. Note that the time scale
to return to a state of large HS distance, $\tau_{rec}=\protect\underset{t>\tau_{dec}}{\min}(\{t:D_{\psi_{1}\psi_{2}}(t)>1-\epsilon\})$
is very different for the left and right columns. Also, the time scale
for reconstructions $\tau_{rec}$ does not depend on the wave packet
width, but the initial decay time, $\tau_{dec}=\protect\underset{t>0}{\min}(\{t:D_{\psi_{1}\psi_{2}}(t)<\overline{D}_{\psi_{1},\psi_{2}}\})$
where $\overline{D}_{\psi_{1},\psi_{2}}:=\lim_{T_{\mathrm{max}}\to\infty}T_{\mathrm{max}}^{-1}\int_{0}^{T_{\mathrm{max}}}D_{\psi_{1},\psi_{2}}(t)dt$
does. a) {[}resp.~b){]} broad initial wave packets with no {[}resp.~with{]}
barriers. c) {[}resp.~d){]} narrow initial wave packets with no {[}resp.~with{]}
barriers. In all of these plots $L_{A}=2$ and $L_{B}=1$ the barriers
are of strength $V_{1}=10^{6}$ and $V_{2}=2.0\times10^{6}$. {}``Narrow''
Gaussians have width $\sigma=0.005$ while {}``broad'' Gaussians
have width $\sigma=0.125$. \label{fig:Time-evolution}}
\end{figure}

In this section we modify the free dynamical evolution by introducing
two delta barriers at the boundaries of region $A$, at positions
$\pm L_{A}/2$, i.e.~we consider the potential $V(x)=V_{1}\delta(x-L_{A}/2)+V_{2}\delta(x+L_{A}/2)$,
where $V_{i}$ measures the strengths of the barriers. Eigenstates
and eigenvalues of this system can be found by integrating Schrödinger's
equation in the neighborhood of the barriers, and imposing continuity
of the wave function. As a result one obtains a transcendental equation
for the quantum number $k$, which is solved using a numerical root
finder based on the bisection method \cite{vetterling_numerical_1992}.
Again, for finite size $L$, the spectrum is purely (countably) discrete.
The initial state is expressed in the Hamiltonian eigenbasis according
to $\psi_{j}(0,x)=\sum_{k}\phi_{k}(x)\langle\phi_{k}|\psi_{j}\rangle$,
keeping as many terms in order to reach a fidelity of at least $0.99$.
The eigenfunctions themselves are just piece-wise continuous combinations
of plane waves in region $A$ and $B$. All this information is then
inserted in Eq.~(\ref{eq:pij_def}) and eventually into Eq.~(\ref{eq:D_Markov}).

Introducing delta barriers of strength $V_{i}$, separating system
and bath, causes several profound modifications of the time evolution,
as observed in Fig.~\ref{fig:Time-evolution} (b) and (d). First,
the time scale for recurrences is increased by several orders of magnitude,
as this is now controlled by the tunneling probabilities through the
barriers. This can be defined as $\tau_{rec}=\underset{t>\tau_{dec}}{\min}(\{t:D_{\psi_{1}\psi_{2}}(t)>1-\epsilon\})$
where $\tau_{dec}=\underset{t>0}{\min}(\{t:D_{\psi_{1}\psi_{2}}(t)<\overline{D_{\psi_{1}\psi_{2}}}\})$
and $\overline{D}_{\psi_{1},\psi_{2}}:=\lim_{T_{\mathrm{max}}\to\infty}T_{\mathrm{max}}^{-1}\int_{0}^{T_{\mathrm{max}}}D_{\psi_{1},\psi_{2}}(t)dt$.
Similarly, the characteristic time scale of the faster oscillations
due to the dispersing wave packets is modified by the tunneling process
through the barriers, albeit in a non-trivial manner. Although the
effective number of states does not seem to be drastically modified,
now a dominant mode is singled out among the $N^{\ast}$ states. This
is clearly visible as an underlying modulation in Fig.~\ref{fig:Time-evolution}
(b) and (d).

\begin{figure}
\begin{centering}
\includegraphics[width=8cm,height=6cm]{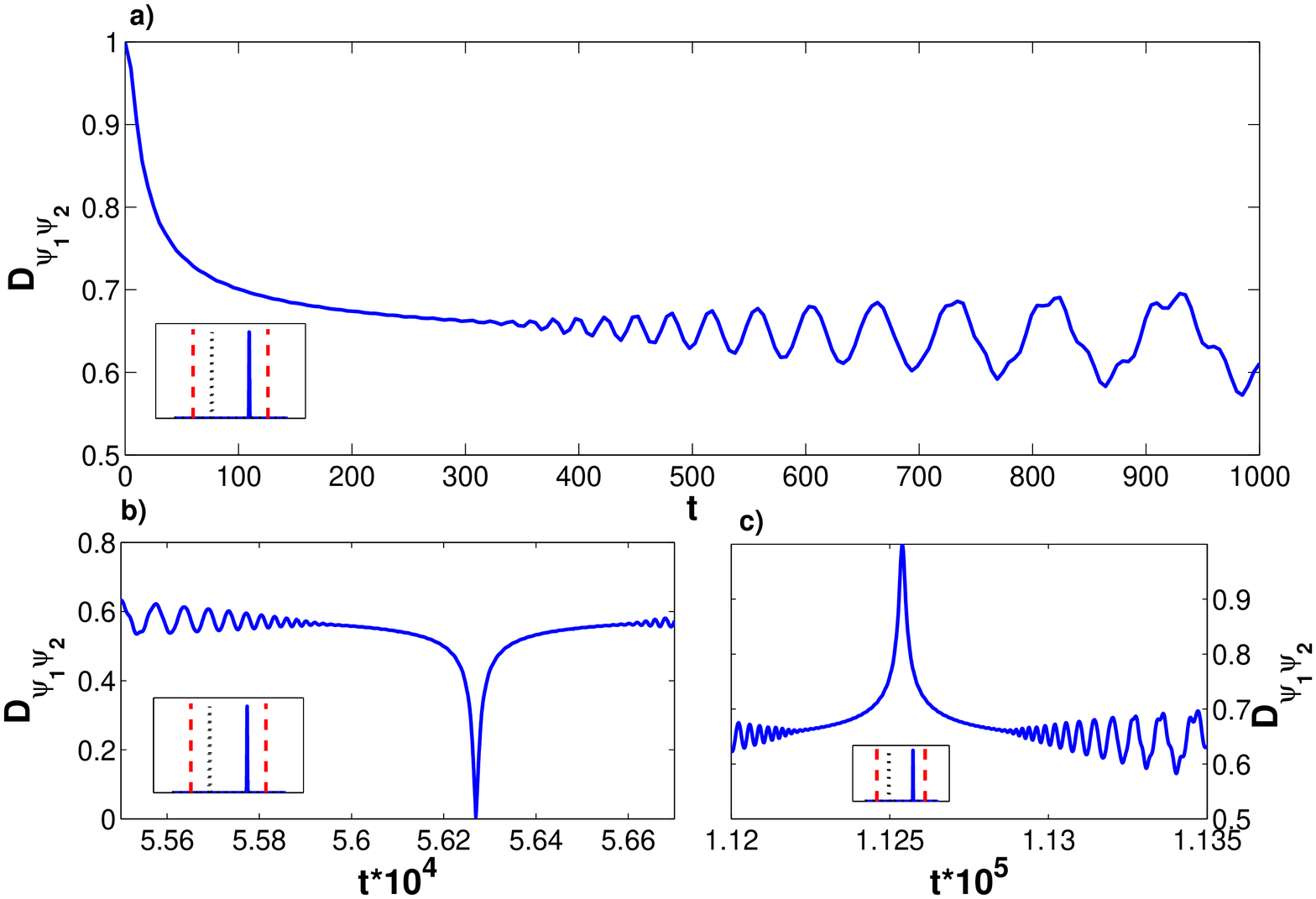} 
\par\end{centering}

\caption{(Color online) Three characteristic features in the time evolution,
all for narrow initial wave packets (indicated in the insets). a)
Initial smooth decay of $D_{\psi_{1},\psi_{2}}(t)$ followed by non-Markovian
oscillations. b) Non-monotonic decay (and subsequent increase) c)
Reconstruction consisting of a smooth increase in the HS distance
back to 1, followed by a smooth decay analogous to (a). These features
re-occur at almost periodically at larger times but do eventually
weaken and disappear. In all of these plots $L_{A}=2$ and $L_{B}=1$
the barriers are of strength $V_{1}=10^{6}$ and $V_{2}=2.0\times10^{6}$
\label{fig:Three-characteristic-features}}
\end{figure}

In addition to these two phenomena, one also observes an extended
pseudo-Markovian initial transient decay in the HS distance, best
seen Fig.~\ref{fig:Three-characteristic-features} (a), where we
zoom into the early-time response. During this transient period, the
two initial wave packets become less distinguishable. The time scale
$t_{NM}$ where non-Markovian oscillations set in will be discussed
later.

In Fig.~\ref{fig:Three-characteristic-features} (b) we observe an
event where $D_{\psi_{1},\psi_{2}}(t)$ distance sharply decays at
some time and the states become almost indistinguishable in region
$A$. We also observe in Fig.~\ref{fig:Three-characteristic-features}
(c) that the initial wave functions become again perfectly distinguishable
(i.e.~orthogonal: $D_{\psi_{1},\psi_{2}}(t)=1$).

\begin{figure}
\begin{centering}
\includegraphics[width=8cm]{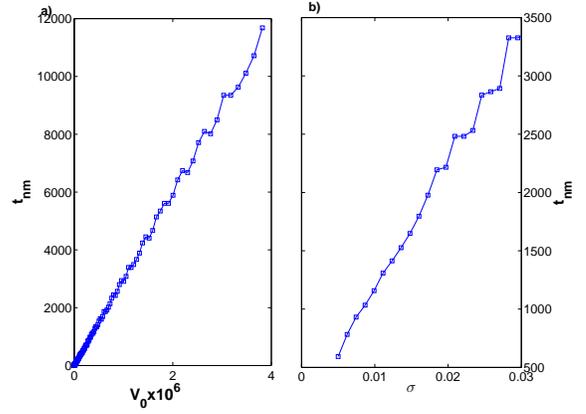} 
\par\end{centering}

\caption{(Color online) Control of the timescale $t_{NM}$ for the onset of
non-Markovian oscillations following the initial transient decay of
the HS distance. a) $t_{NM}$ versus barrier strength $V_{0}$ ($V_{1}=V_{0}$
and $V_{2}=2V_{0}$). b) $t_{NM}$ versus the width of the initial
wave packets, $\sigma$. The dependence is approximately linear in
both cases. $t_{NM}$ is operationally defined as the time when the
HS distance $D_{\psi_{1},\psi_{2}}(t)$ increases by 10\% from its
minimum value before $t_{NM}$. Main plots are on a log-log scale,
whereas insets are on a linear scale.\label{fig:quasi-Markovian}}
\end{figure}

The pseudo-Markovian event observed in Fig.~\ref{fig:Three-characteristic-features}
clearly defines a characteristic time scale $t_{NM}$ which is the
transient for which the system behaves in a Markovian fashion before
non-Markovian oscillations occur %
\footnote{Operationally $t_{NM}$ is the first occurrence of time $t$ for which
$\sigma(t,\Gamma_{1},\Gamma_{2})>0$%
}. We have operationally defined $t_{NM}$ as the time when the HS
distance $D_{\psi_{1},\psi_{2}}(t)$ increases by 10\% from its minimum
value before $t_{NM}$. We have verified that this time constant $t_{NM}$
scales linearly with the strength of the barriers $V_{0}$, (for the
case $V_{1}=V_{0}$ and $V_{2}=2V_{0}$) , and with the width of the
initial Gaussian wave packets, $\sigma$ (see Fig.~\ref{fig:quasi-Markovian}).
Indeed, stronger barriers mean longer tunneling times, and therefore
a longer time for the particles to return from the bath (Fig.~\ref{fig:quasi-Markovian}
(a)). On the other hand narrower wave packets lead to higher occupancy
of the high-energy modes which can more easily pass through the barriers
(Fig.~\ref{fig:quasi-Markovian} (b)).

It can be argued that the dips and peaks observed in Figs.~\ref{fig:Three-characteristic-features}
(b) and (c) are artifacts of a very fine-tuned system-bath aspect
ratio. Let us therefore examine more generic situations with variable
bath size. This will also allow us to explore the level of control
we can exert on the system by tuning feedback effects due to variable
bath size, as well as examine the crossover to a more conventional
notion of bath in the limit $L_{B}\rightarrow\infty$.

\begin{figure}
\begin{centering}
\includegraphics[width=9cm]{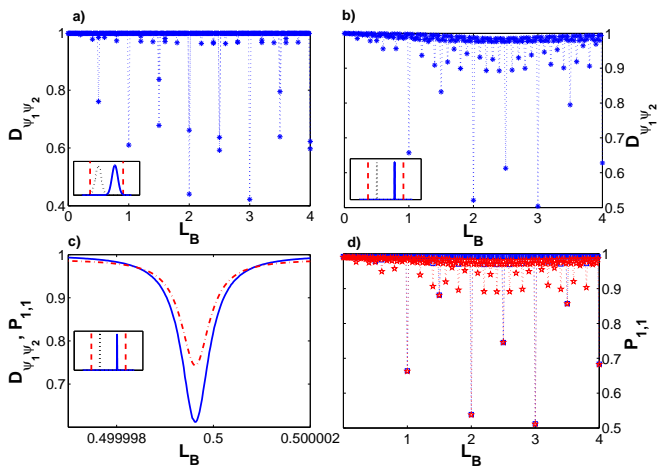} 
\par\end{centering}

\caption{(Color online) Time averaged quantities as a function of bath size
$L_{B}$ for fixed system size $L_{A}=2$. a) $\overline{D}_{\psi_{1},\psi_{2}}$
versus $L_{B}$ for broad initial Gaussians: narrow dips in trace
distance occur at lengths for which a significant portion of modes
are able to tunnel into the bath. b) Same as (a), but with initially
narrow Gaussians. c) A zoom on the peak seen in (b) at $L_{B}=1$.
Note that $\overline{D}_{\psi_{1},\psi_{2}}$ increases smoothly as
$L_{B}$ is moved away from $1/2$. This plot also shows the time
average of $p_{1,1}$ as a (red) dashed line. d) Time average of $p_{1,1}$versus
$L_{B}$ for narrow Gaussians, (red) stars, and broad Gaussians, (blue)
squares. Note the similarity between these plots and plots (a) and
(b). In all of these plots barriers are of strength $V_{1}=10^{6}$
and $V_{2}=2.0\times10^{6}$, {}``narrow'' Gaussians have width
$\sigma=0.005$ whereas {}``broad'' Gaussians have width $\sigma=0.125$.
\label{fig:Time-averaged}}
\end{figure}

The fine tuning required to control the tunneling can be seen in Fig.~\ref{fig:Time-averaged}
(a)-(d). Fig.~\ref{fig:Time-averaged} (a) shows the time averaged
HS distance $\overline{D}_{\psi_{1},\psi_{2}}:=\lim_{T_{\mathrm{max}}\to\infty}T_{\mathrm{max}}^{-1}\int_{0}^{T_{\mathrm{max}}}D_{\psi_{1},\psi_{2}}(t)dt$
versus size of the bath. One observes that only at very specific values
of $L_{B}$ a significant deviation from perfect distinguishability
is achieved. Comparison to Fig.~\ref{fig:Time-averaged} (d) which
shows average probability for one of the particles to be found in
the system, indicates that the underlying cause of this lack of distinguishability
is that the wave function cannot significantly tunnel out of the system
for most values of $L_{B}$. We notice that in the case of narrow
initial Gaussians (Fig.~\ref{fig:Time-averaged} (b)) the particle
is able to tunnel out of the system for more values of $L_{B}$, as
more eigenmodes of the system have significant amplitudes, but otherwise
this result is analogous to the one shown in Fig.~\ref{fig:Time-averaged}
(a).

The reason for the behavior seen in Fig.~\ref{fig:Time-averaged}
relates to the fact that the energy scale of all eigenmodes of the
system considered here is much smaller than the barrier strengths.
For this reason, the only modes which are allowed to have appreciable
amplitudes both in the system and the bath are those where both of
the barriers are very close to a node.

Fig.~\ref{fig:Time-averaged} (c) shows that both the average HS
distance and the average probability to be in the bath can be controlled
by fine tuning $L_{B}$. As the ratio of $L_{B}$ vs.~$L$ is moved
further from away from $1/2$, the modes become separated into those
which are isolated in the system, and those which are isolated in
the bath. As this happens, tunneling is reduced, and consequently
the particles become more distinguishable on average.

\begin{figure}
\begin{centering}
\includegraphics[width=6cm,height=4cm]{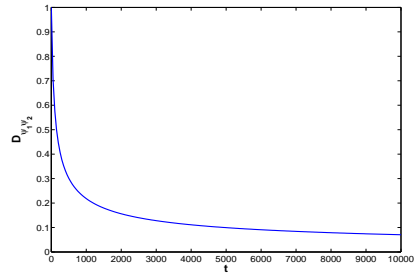} 
\par\end{centering}

\caption{(Color online) Monotonic ({}``Markovian'') decay of the HS distance
versus time for a pair of initially narrow Gaussians with $L_{B}\rightarrow\infty$,
$L_{A}=1$ and $V_{1}=V_{2}=V_{0}=10^{6}$. \label{fig:escape_dist}}
\end{figure}

We now turn our interest to the case where $L_{A}$ is fixed and $L_{B}\rightarrow\infty$.
Here we set the two barrier strengths equal for simplicity. In this
case the particles do not return after escaping the system, so the
evolution is Markovian. Fig.~\ref{fig:escape_dist} shows the resulting
trace distance for the case of a pair of initially narrow Gaussians.
As expected this plot indicates Markovian decay as the wave functions
escape the system.

\section{Conclusions}

In summary, we have investigated non-Markovian effects arising from
a tunable finite bath. These can be quantified by a measure that tracks
particles flow out of and into a system that is connected to the bath.
We have identified various time scales for rephasing events that depend
non-trivially on the bath size, on the tunneling barrier potentials
between the system and the bath, and on the shape of the wave functions
with which the evolution is initiated. In particular, we found that
substantial rephasing can be achieved by fine tuning the bath length
and by choosing initial states with significant high frequency components,
allowing the wave packets to tunnel efficiently between the system
and the bath. One can envision physical realizations of such a setup
in the context of nanoelectronics and nanophotonics. For example,
a photonic microcavity can act as the system, connected via semi-transparent
mirrors to an external cavity that acts as the bath. Then the transparency
of the mirrors corresponds to the barrier potential, and the length
of the external cavity sets the time scale for major rephasing events.
An implementation in the context of nanoelectronics may be even more
interesting, because in this case the effects of electron interactions
on non-Markovian system dynamics could be studied as well.

\paragraph*{Acknowledgements:}

The authors would like to thank Tameem Albash and Paolo Zanardi for
useful discussions. The numerical computations were carried out on
the University of Southern California high performance supercomputer
cluster. This research has been supported by the ARO MURI grant W911NF-11-1-0268.
S.~Haas would like to thank the Humboldt Foundation for support. 

\bibliographystyle{unsrt}
\bibliography{non-Markovian_stephan}

\end{document}